\documentclass{iau}
\usepackage{graphicx}

\newcommand{\apj}{{\it ApJ}}                                                                                           
\newcommand{\aj}{{\it AJ}}                                                                                             
\newcommand{\mnras}{{\it MNRAS}}                                                                                       
\newcommand{\aanda}{{\it A\&A}}                                                                                                                                                                                 

\title[IAUS289.~~Advancing the Physics of Cosmic Distances]{Advancing
  the Physics of Cosmic Distances: Conference Summary}

\author[Richard de Grijs]   
{Richard de Grijs}

\affiliation{Kavli Institute for Astronomy and Astrophysics, Peking
  University, Yi He Yuan Lu 5, Hai Dian District, Beijing 100871,
  China \\ email: {\tt grijs@pku.edu.cn}}

\pubyear{2012}
\volume{289}  
\pagerange{xxx--yyy}
\jname{Advancing the Physics of Cosmic Distances}
\editors{Richard de Grijs \& Giuseppe Bono, eds.}
\begin{document}

\maketitle

\begin{abstract}
Knowing the distance of an astrophysical object is key to
understanding it. However, at present, comparisons of theory and
observations are hampered by precision (or lack thereof) in distance
measurements or estimates. Putting the many recent results and new
developments into the broader context of the physics driving cosmic
distance determination is the next logical step, which will benefit
from the combined efforts of theorists, observers and modellers
working on a large variety of spatial scales, and spanning a wide
range of expertise. IAU Symposium 289 addressed the physics underlying
methods of distance determination across the Universe, exploring the
various approaches employed to define the milestones along the
road. The meeting provided an exciting snapshot of the field of
distance measurement, offering not only up-to-date results and a
cutting-edge account of recent progress, but also full discussion of
the pitfalls encountered and the uncertainties that remain. One of the
meeting's main aims was to provide a roadmap for future efforts in
this field, both theoretically and observationally.
\end{abstract}




\firstsection 
\section{Introduction}

Knowing the distance of an astrophysical object is key to
understanding it: without an accurate distance we do not know how
bright it is, how large it is, or even (for long distances) when it
existed. But astronomical distance measurement is a challenging task,
since the only information we have about any object beyond our solar
system is its position (perhaps as a function of time) and its
brightness (as a function of wavelength and time).

In 1997, the {\sl Hipparcos} space mission provided (for the first
time) a significant number of absolute trigonometric parallaxes at
milliarcsec (mas)-level precision across the whole sky, which had a
major impact on all fields of astrophysics. In addition, during the
past ten years, the use of ground-based 8--10 m-class optical and
near-IR telescopes ({\sl Keck, VLT, Gemini, Subaru}) and space
observatories (the {\sl Hubble Space Telescope [HST], Spitzer,
  Herschel, Chandra, XMM-Newton}) have provided an unprecedented
wealth of accurate photometric and spectroscopic data for stars and
galaxies in the local Universe. Radio observations, particularly with
the Very Long Baseline Array ({\sl VLBA}) and the Japanese {\sl VERA}
array, have achieved 10 micro-arsecond astrometric accuracy. Moreover,
stellar models and numerical simulations are now providing accurate
predictions of a broad range of physical phenomena, which can in
principle be tested using accurate spectroscopic and astrometric
observations. However, at present, comparisons of theory and
observations are mainly hampered by precision (or lack thereof) in
distance measurements/estimates.

IAU Symposium 289 highlighted the tremendous amount of recent and
continuing research into a myriad of exciting and promising aspects of
accurately pinning down the cosmic distance scale. Putting the many
recent results and new developments into the broader context of the
physics driving cosmic distance determination is the next logical
step, which will benefit from the combined efforts of theorists,
observers and modellers working on a large variety of spatial scales,
and spanning a wide range of expertise.

This is a very exciting time in the context of this Symposium. Very
Long Baseline Interferometry (VLBI) sensitivity is being expanded
allowing, for example, direct measurement of distances throughout the
Milky Way and even slightly beyond Local Group galaxies. The field
will benefit from expert input to move forward into the era of {\sl
  Gaia}, optical-interferometer and Extremely Large Telescope-driven
science, which (for example) will allow us to determine Coma-cluster
distances without having to rely on secondary distance indicators,
thus finally making the leap to accurate distance measurements well
beyond the Local Group of galaxies.

In this Symposium, we managed to bring together experts on various
aspects of distance determinations and (most importantly) the
underlying physics enabling this (without being restrictive in areas
where statistical and observational approaches are more relevant),
from the solar neighbourhood to the edge of the Universe, exploring on
the way the various methods employed to define the milestones along
the road. We aimed at emphasising, where possible, the physical bases
of the methods and recent advances made to further our physical
insights. We thus aimed to provide a snapshot of the field of distance
measurement, offering not only up-to-date results and a cutting-edge
account of recent progress, but also full discussion of the pitfalls
encountered and the uncertainties that remain. We ultimately aimed to
provide a roadmap for future efforts in this field, both theoretically
and observationally.

Although our focus was on techniques of distance determination, this
is intimately linked to many other aspects of astrophysics and
cosmology. On our journey from the solar neighbourhood to the edge of
the Universe, we encountered stars of all types, alone, in pairs and
in clusters, their life cycles, and their explosive ends: binary
stars, in particular, play an important role in this context, e.g. in
pinning down accurate distances to the Pleiades open cluster and Local
Group galaxies, as well as in future ground- and space-based surveys;
the stellar content, dynamics, and evolution of galaxies and groups of
galaxies; the gravitational bending of starlight; and the expansion,
geometry and history of the Universe. As a result, the Symposium
offered not only a comprehensive study of distance measurement, but a
tour of many recent and exciting advances in astrophysics.

\section{Key areas of new and/or sustained progress}

In 16 reviews, 16 invited and 35 contributed talks, as well as more
than 50 high-quality posters, the Symposium presented a venue for
lively debate, exciting new results and a number of potentially
ground-breaking new announcements.

\subsection{Has the Pleiades distance controversy finally been resolved?}

The Pleiades open cluster is a crucial rung of the local distance
ladder, whose calibration affects many fundamental aspects of stellar
astrophysics. However, the original {\sl Hipparcos} parallaxes
(Mermilliod et al. 1997; van Leeuwen \& Hansen-Ruiz 1997; Robichon et
al. 1999; van Leeuwen 1999), as well as the recalibrated astrometry
(van Leeuwen 2007a,b), yielded distances to the individual member
stars and the open cluster as a whole that were systematically lower
than those resulting from previous ground-based distance
determinations. The latter were predominantly based on the
main-sequence fitting technique, because prior to the successful {\sl
  Hipparcos} mission stellar parallaxes at the distance of the
Pleiades were too small to be measurable reliably with contemporary
instrumentation.

Doubt was initially cast on the original {\sl Hipparcos} analysis,
which required advanced mathematical techniques to solve
simultaneously for the positions, motions and distances of 118,000
stars. Although the {\sl Hipparcos} recalibration reduced the
discrepany slightly, the difference remains too large for comfort: the
variation in distance modulus implied is approximately 0.2--0.3 mag
(Pinsonneault et al. 1998), while the difference in parallax required
is of order 1 mas, but note that the absolute uncertainty in {\sl
  Hipparcos} parallaxes is only 0.1 mas (Arenou et al. 1995; Lindegren
1995).

The controversy has, thus, not been fully resolved, and all methods
applied to date are affected by their own unique sets of uncertainties
(see e.g. Valls-Gabaud 2007). To account for the {\sl Hipparcos}
distance, stellar models would require changes in physics or input
parameters that are too radical to be reasonable, e.g. changes in the
Pleiades' characteristic metallicity or helium abundance, or in the
mixing length, an age differential between local and Pleiades member
stars or an unusual spatial distribution (i.e. depth effects). Most
models applied to resolve the Pleiades controversy include many
assumptions and simplifications which may well dominate or negate the
need for the proposed small evolutionary correction between the
Pleiades and local stars, e.g. in terms of stellar structure
(rotation, convection, magnetic fields), stellar evolution and stellar
atmospheres. VLBI observations may come to the rescue in this context:
1\% parallax precision for individual Pleiades stars is anticipated
(0.5\% for objects in Gould's Belt). The jury is thus still out on the
final resolution of the Pleiades controversy, but there appears to be
light at the end of the tunnel.

\subsection{Further refinements of the distance to the Galactic Centre}

The exact distance from the Sun to the Galactic Centre, R$_0$, serves
as a benchmark for a variety of methods used for distance
determination, both inside and beyond the Milky Way. Many parameters
of Galactic objects, such as their distances, masses and luminosities,
and even the Milky Way's mass and luminosity as a whole, are directly
related to R$_0$. Most luminosity and many mass estimates scale as the
square of the distance to a given object, while masses based on total
densities or orbit modelling scale as distance cubed. This dependence
sometimes involves adoption of a rotation model of the Milky Way, for
which we also need to know the Sun's circular velocity with high
accuracy.

Significant efforts have been expended in recent years to reduce the
uncertainties in and narrow down the actual distance to the Galactic
Centre, using a large variety of mostly independent methods. Detailed
orbit modelling of the so-called S stars orbiting Sagittarius A*
(believed to be almost coincident with the supermassive black hole in
the Galactic Centre) yields R$_0 = 8.20 \pm 0.15$ (statistical) $\pm
0.31$ (systematic) kpc (Gillessen et al., in prep.) or R$_0 = 7.7 \pm
0.4$ kpc (Morris et al. 2012), depending on one's assumptions about
the central black hole mass and the associated uncertainties. We will
need to wait until at least 2019, when we will finally have
high-accuracy direct astrometric measurements of a full orbit of star
S2, for significantly reduced errors in these distance estimates.  The
current accuracy of R$_0$ determinations based on orbit modelling
compares well with the results from, e.g. Cepheid-based distances.
Majaess (2010), using the {\sc ogle} (Optical Gravitational Lensing
Experiment) fields, finds R$_0 = 8.1 \pm 0.6$ kpc, while Dambis (2009)
reported R$_0 = 7.58 \pm 0.40$ kpc. An alternative distance tracer in
the form of Mira variables results in R$_0 = 8.24 \pm 0.08$ (stat.)
$\pm 0.43$ (syst.) kpc (Matsunaga et al. 2009), based on a sample of
143 Miras. Thus, although the exact value of R$_0$ remains open to
debate, it is clear that the IAU-recommended value of 8.5 kpc is too
large, but we are not sure whether the true value should be greater or
less than 8.0 kpc

In terms of refining our knowledge of the Galactic rotation
parameters, significant progress has also been made in recent
years. Based on 5000 hours of VLBA observations, the {\sc BeSSeL}
survey obtained parallaxes and proper motions of $>400$ sources. The
{\sc BeSSeL} team reports a new Galactic rotation velocity of
$\Theta_0 = 243 \pm 7$ km s$^{-1}$ (for R$_0 = 8.38 \pm 0.18$ kpc) or
$\Theta_0 = 236 \pm 10$ km s$^{-1}$ for R$_0 = 8.2 \pm 0.3$ kpc based
on stellar orbits and proper motions of Sgr A*. The Japanese VERA
team, meanwhile, obtained parallaxes of 30 objects, resulting in R$_0
= 8.05 \pm 0.45$ kpc and $\Omega_0 \equiv (\Theta_0/R_0) = 31.09 \pm
0.78$ km s$^{-1}$ kpc$^{-1}$.

\subsection{Stellar distance tracers: calibration of Local Group distances}

A significant fraction of the meeting was devoted to discussions about
the use, reliability and calibration of pulsating stars as distance
tracers, essentially based on using their period--luminosity
relations. This is an extensive field, in which much progress has been
made since the Cepheid period--luminosity relation was first
established by Henrietta Leavitt a century ago.

Much of the current debate centres on whether or not the relation for
Cepheids exhibits a single slope or is perhaps better defined by two
segments with independently determined slopes. It appears that
observations at longer wavelengths, particularly in the near- and
mid-IR, may bring closure to this issue. In addition to unequivocally
yielding single-slope relations, the associated error bars are much
reduced, hence leading to distance estimates affected by significantly
reduced uncertainties compared to the use of optical
period--luminosity(--colour) relations. Reddening corrections remain
among the key sources of uncertainty. Additional sources of
uncertainty include the alleged effects of circumstellar envelope
variability, a source of error that has long been overlooked and
neglected, and the maximum useful period for Cepheid
period--luminosity relation applications (ultralong-period Cepheids do
not seem to obey a clear-cut relationship of this type).

It was suggested that red-supergiant Mira variables may be better
distance tracers than Cepheids under certain circumstances, given that
they are brighter and associated with old(er) stellar populations. As
massive stars, Cepheids are by definition confined to young stellar
populations. Ideally, linking up both tracers in the same galaxy will
conclusively constrain the distance debate.

Among the brightest non-variable distance tracers, recent years have
seen significant improvements in the accuracy of using stars at the
tip of the red giant branch. But all these methods rely on secondary
calibration, i.e. on the presumption that we understand the underlying
physics of both nearby and distant objects of the same type. Geometric
distance methods out to the Local Group galaxies are, unfortunately,
few and far between. It was therefore encouraging to note that
eclipsing binary systems have been detected and used to constrain the
distance to IC 1613 with encouraging accuracy. This may lead to IC
1613 eventually being designated as a new Local Group distance
benchmark.

\subsection{LMC distance: the first step of the extragalactic distance ladder}

The Magellanic Clouds, and in particular the Large Magellanic Cloud
(LMC), represent the first rung on the extragalactic distance
ladder. The galaxy hosts statistically large samples of potential
standard candles, including many types of variable stars. They are all
conveniently located at roughly the same distance -- although for
detailed distance calibration the LMC's line-of-sight depth and 3D
morphology must also be taken into account -- and relatively
unaffected by foreground extinction. The LMC's unique location allows
us to compare and, thus, cross correlate and calibrate a variety of
largely independent distance indicators, which can, in turn, be
applied to more distant targets.

The distance to the LMC has played an important role in constraining
the value of the Hubble constant. The {\sl HST} Key Project on the
extragalactic distance scale (Freedman et al. 2001) used a revised
calibration of the Cepheid period--luminosity relation (adopting the
maser-based distance to NGC 4258) and numerous secondary techniques to
obtain a distance modulus to the LMC of $(m-M)_0 = 18.50 \pm 0.10$ mag
-- corresponding to a distance $D_{\rm LMC} = 50.1^{+1.4}_{-1.2}$ kpc
-- and H$_0 = 72 \pm 3$ (stat.) $\pm 7$ (syst.) km s$^{-1}$
Mpc$^{-1}$. Trends in subsequent LMC distance determinations have been
questioned by Schaefer (2008): he argued that all 31 measurements
published between 2001 and early 2008 cluster too tightly around the
{\sl HST} Key Project's value and suggested that this may imply a
`bandwagon effect', i.e. publication bias.

Once again, going to near- and mid-IR wavelengths may enable us to
reduce the uncertainties in the distance to the LMC. At present,
2--3\% distance accuracy is already achievable, and this may be
improved to $\sim 1$\% in the near future! For instance, the Carnegie
Hubble Program, using data from the warm {\sl Spitzer} mission,
derived $(m-M)_0 = 18.477 \pm 0.034$ mag (Freedman et al. 2012), while
Ripepi et al. (2012) used {\sl VISTA} observations to arrive at
$(m-M)_0 = 18.46 \pm 0.03$ mag. These distances are comfortably close
to and within the mutual uncertainties of the direct, geometric
distance determination based on eclipsing binaries by Pietrzy\'nski et
al. (2012), $(m-M)_0 = 18.48 \pm 0.01$ (stat.)  $\pm 0.04$ (syst.)
mag.

\subsection{Rotational parallaxes/water masers: new standard benchmarks?}

The technique of VLBI is not only useful in the context of resolving
the distance to the Pleiades, it is also increasingly used to measure
extragalactic proper motions. In turn, this enables geometric distance
determination out to some 100 Mpc, including to the nearby galaxies
NGC 4258, M33, UGC 3789, NGC 6264, and many more. Combined with {\it a
  priori} information on a galaxy's inclination with respect to our
line of sight and its rotation curve, based on radial velocity
measurements, we can construct an accurate, slightly warped
`tilted-ring' model of the galaxy's dynamical structure, usually
assuming circular orbits (although this assumption does not result in
major systematic uncertainties).  This, in turn, allows correlation of
the angular proper motion measurements with the rotational velocity
information obtained in linear units and, thus, provides an
independent distance measurement.

Much has been made of the original application of water maser
measurements in NGC 4258, but in the mean time this technique has been
extended to other nearby systems. Initial efforts to determine the
distance to the Local Group galaxy M33 have thus far resulted in
$D_{\rm M33} = 750 \pm 140 \pm 50$ kpc, where the first uncertainty is
related to uncertainties in the H{\sc i} rotation model adopted for
the galaxy, and the second uncertainty comes from the proper motion
measurements. The meeting was told that 10\% distance accuracy will be
achievable eventually, provided that the team retains access to the
Green Bank Telescope, which was recently slated for closure because of
severe funding constraints. The prospects for application of this
technique to M31 are moderately positive, although at present only two
water masers have been identified in the galaxy that are potentially
useful; meanwhile, Cepheid variables enable a distance estimate of
$D_{\rm M31} = 752 \pm 27$ kpc (3\%). The {\sl Square Kilometre Array
  (SKA)} may have a major role to play in this context

Simultaneously, the Megamaser Cosmology Project aims at using
extragalactic maser sources to direcyly measure H$_0$ in the Hubble
flow, which is clearly a very challenging endeavour at distances
$>100$ Mpc! Their preliminary results look promising however: using
NGC 6264 ($D = 137$ Mpc) as a benchmark, they find H$_0 = 74 \pm 10$
km s$^{-1}$ Mpc$^{-1}$.

\subsection{Nearby galaxy samples}

Beyond the distances where common geometric methods or even Cepheid
period--lumi\-nos\-ity relations offer a way to determine reasonably
accurate distances, many secondary methods of distance determination
have been developed. In almost all cases, their reliability depends on
a proper understanding of the underlying physics. And this is where
some of the key remaining problems originate, leading to `annoying'
uncertainties that are hard to reduce.

For instance, the planetary nebula luminosity function (PNLF) has long
been used as a distance indicator. In the narrow-band filter centred
on the [O{\sc iii}]$\lambda$5007{\AA} emission line, the PNLF in
nearby galaxies is defined by a universal, sharp cut-off at the bright
end. This feature can be used as a standard candle, resulting in
distances to the planetary nebulae's host galaxies accurate to $\sim
25$\% or better. However, at the meeting we learnt that this sharp
cut-off may, in fact, not be so sharp, particularly in giant
elliptical galaxies. This implies that we really need to improve our
physical understanding of the processes that lead to the establishment
of the PNLF. (A serious problem in this context is that worldwide
there is a trend leading to a general loss of narrow-band capabilities
at major research observatories!)

Secondly, the often used Tully--Fisher relation, which relates a
galaxy's luminosity to its rotational velocity, enables us to
determine highly accurate distance, but one should realise that its
use is actually based on numerous simplifications, assumptions and
degeneracies (e.g. on the assumption of an asymptotic value of the
rotation curve, adoption of halo mass scaling relations), but it works
somehow, and surprisingly well! Application of this technique yields
values of H$_0$ in the same ballpark as those obtained from Cepheids,
with uncertainties of order 10\%. Expressed in worldly units, we were
told that one can obtain relevant observations for large samples of
galaxies at a cost ranging from from US\$ 200 to US\$ 15,000 per
galaxy.

A promising alternative method of distance determination out to galaxy
clusters in the Hubble flow is found in the technique of surface
brightness fluctuations (SBFs). The relative SBF distance to the Coma
cluster with respect tot the Virgo and Fornax clusters is well
determined: $(m-M)_{0,\rm Coma} = 34.98 \pm 0.06/34.96 \pm 0.07$ mag,
or $\Delta (m-M)_0 = 3.89 \pm 0.06$ mag relative to the Virgo
cluster. This leads to a robust distance estimate to the Coma cluster
of $D_{\rm Coma} = 99 \pm 3$ Mpc.

\subsection{In the Hubble flow and beyond}

Type Ia supernovae (SNe Ia) remain excellent distance indicators out
to moderate redshifts, despite many lingering uncertainties as regards
the underlying physics, including those related to the apparent
progenitor diversity. However, we are approaching systematic limits
hampering efforts to further reduce the associated
uncertainties. These systematic effects include, e.g. the precision of
photometric calibrations and the fairly limited numbers of
observations of SNe Ia.

Current values of H$_0$ based on measuring gravitational-lens time
delays range from approximately 50 to 85 km s$^{-1}$ Mpc$^{-1}$. Most
of the uncertainties originate from sometimes poorly constrained model
assumptions, e.g. adopton of isothermal profiles, sometimes in the
presence of external tidal fields, the central concentration
degeneracy, environmental density distributions or multiple
lenses. One particularly interesting quadruple gravitational lens,
B1608+656, long held the distinction of allowing the most
straightforward application of time delay measurements. The value of
H$_0$ resulting from these observations was H$_0 = 70.6 \pm 3.1$ km
s$^{-1}$ Mpc$^{-1}$.  Derivation of this result was all but
straightforward, however. It required adoption of the currently
favoured cosmological parameters, {\sl HST} pixel-by-pixel photometry,
{\sl Keck} low-resolution spectroscopy, cosmological $N$-body
simulations, and assumption of a proper, extended source intensity
distribution. Combined with the {\sl Wilkinson Microwave Anisotropy
  Probe (WMAP)} five-year results and assuming a flat geometry, H$_0 =
69.7^{+4.9}_{-5.0}$ km s$^{-1}$ Mpc$^{-1}$ and equation of state, $w =
–0.94^{+0.17}_{-0.19}$ (68\% confidence; Suyu et al. 2010).

More recently, Suyu et al. (2012) improved these cosmological
parameter determinations using a second, well-understood lens,
RXJ1131--1231. The current-best cosmological parameters resulting from
gravitational-lens time-delay measurements are H$_0 =
75.2^{+4.4}_{-4.2}$ km s$^{-1}$ Mpc$^{-1}$ and $w =
-1.14^{+0.17}_{-0.20}$.

On scales of (and distances to) distant galaxy clusters, the
Sunyaev--Zel'dovich effect offers a suitable handle on determinations
of H$_0$. Current determinations depend on the assumptions adopted;
they include H$_0 = 76.9^{+3.9}_{-3.4}$ (stat.) $^{+10.0}_{-8.0}$
(syst.) versus H$_0 = 73.7^{+4.6}_{-3.8} {\strut}^{+9.5}_{-7.6}$ km
s$^{-1}$ Mpc$^{-1}$, adopting hydrostatic equilibrium versus an
isothermal cluster model (Bonamente et al. 2006), H$_0 =
77.6^{+4.8}_{-4.3} {\strut}^{+10.1}_{-8.2}$ km s$^{-1}$ Mpc$^{-1}$ if
one attempts to avoid the cool cluster cores, and H$_0 =
73.2^{+4.3}_{-3.7}$ and H$_0 = 71.4^{+4.4}_{-3.4}$ km s$^{-1}$
Mpc$^{-1}$ for a standard $\Lambda$CDM cosmology versus a flat
Universe with constant $w$ (Holanda et al. 2010).

Other H$_0$ values determined in recent years using up-to-date
techniques and competitive data sets include measurements done as part
of the Carnegie Hubble Program: H$_0 = 74.3 \pm 0.4$ (stat) $\pm 2.1$
(syst.) km s$^{-1}$ Mpc$^{-1}$ (based on calibration in the Local
Group) and H$_0 = 74.3 \pm 2.6 \pm 3.5$ km s$^{-1}$ Mpc$^{-1}$ using
$>400$ galaxies based on mid-IR ($3.6 \mu$m) data. In this regard,
observations with the {\sl James Webb Space Telescope (JWST)} offer
exciting prospects to reach unprecedented 1\% accuracy!

The 6dF Galaxy Survey of local, low-$z$ baryon acoustic oscillations
(BAOs) yields H$_0 = 67 \pm 3.2$ km s$^{-1}$ Mpc$^{-1}$, a 5\% result,
with exciting prospects for further improvement, to approximately
1\%. More importantly, BAO distances are comparable to SNe Ia
distances, which implies that this would allow a crucial cross check
of results. On large scales, {\sl WMAP} has made significant
contributions; based on seven years of observations ({\sl WMAP}-7),
H$_0 = 70.4 \pm 2.5$ km s$^{-1}$ Mpc$^{-1}$. Note that these results
rely on many priors and are not independent. In particular, they
assume a flat geometry, but the value of H$_0$ is degenerate with the
total curvature of the Universe, $k$. If these constraints are
relaxed, H$_0 \mbox{(WMAP-7)} = 53^{+15}_{-13}$ km s$^{-1}$
Mpc$^{-1}$, although the meeting was also told that adoption of
Modified Newtonian Dynamics would lead to lower values of H$_0$ too,
by $\sim 20$\%.

\section{Concluding thoughts}

\begin{figure*}[h!]
\begin{center}
\includegraphics[width=11.5cm]{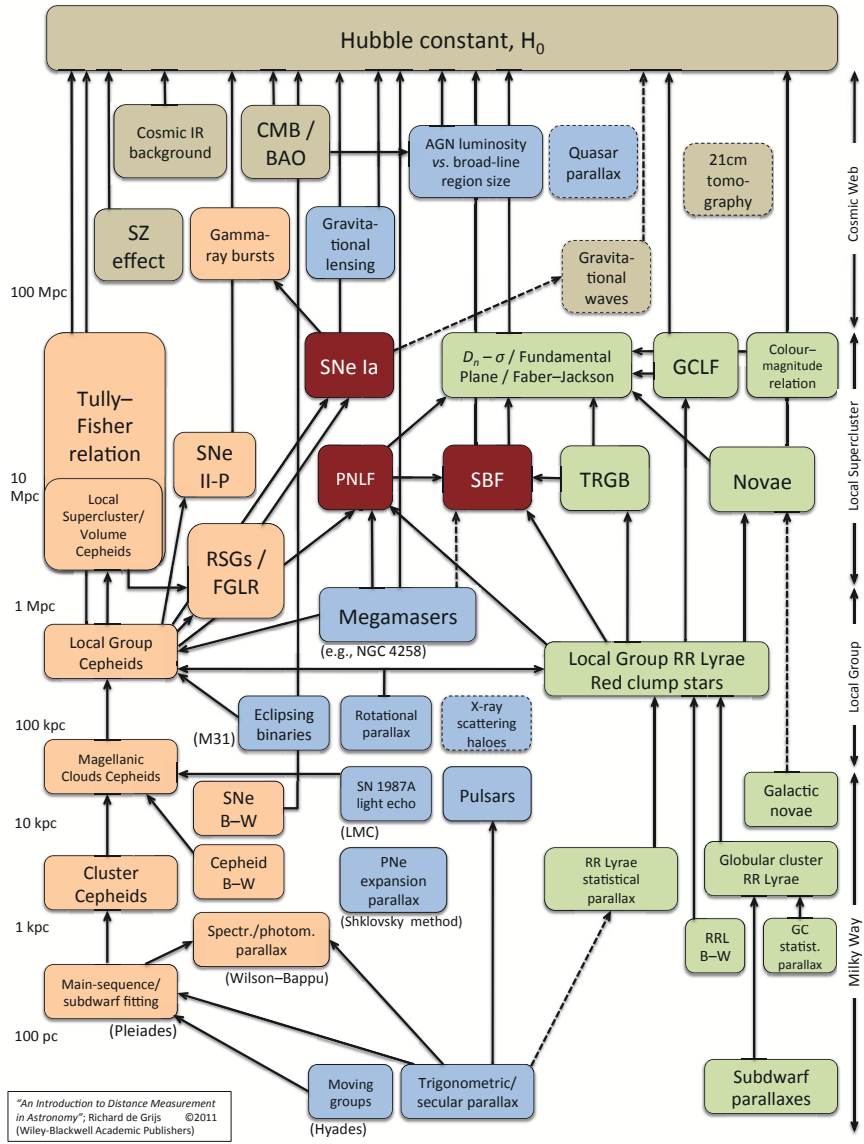}
\end{center}
\caption{Updated, present-day distance ladder, based on an original
  idea by Ciardullo (2006). Light orange: Methods of distance
  determination associated with active star formation (`Population I',
  intermediate- and high-mass stars). Light green: Distance tracers
  associated with `Population II' objects/low-mass stars. Blue:
  Geometric methods. Red: Supernovae (SNe) Ia, the planetary nebulae
  (PNe) luminosity function (PNLF) and surface-brightness fluctuations
  (SBF) are applicable for use with both Populations I and II. Light
  brown: Methods of distance or H$_0$ determination which are not
  immediately linked to a specific stellar population. Dashed boxes:
  Proposed methods. Solid, dashed arrows: Reasonably robust, poorly
  established calibrations. B--W: Baade--Wesselink. RRL: RR
  Lyrae. RSGs/FGLR: Red supergiants/flux-weighted gravity--luminosity
  relationship. TRGB: Tip of the red-giant branch. GCLF: Globular
  cluster (GC) luminosity function. SZ: Sunyaev--Zel'dovich. CMB/BAO:
  Cosmic microwave background/baryon acoustic
  oscillations. Colour--magnitude relation: Refers to galactic colours
  and magnitudes. {\it (adapted from de Grijs 2011)}}
\label{distanceladder.fig}
\end{figure*}

Significant recent progress has been achieved in establishing an
increasingly firm and robust distance ladder, where possible based on
well-understood physics. Nevertheless, uncertainties -- both
systematic and statistical -- persist, even for the nearest and
presumably best understood rungs of the distance ladder, resulting
from different observational or technical approaches, as well as from
our incomplete theoretical understanding of relevant physical
aspects. An example of such lingering systematic uncertainties and the
associated controversy is related to the role of the Pleiades open
cluster as a crucial nearby rung of the cosmic distance
ladder. Reconciliation of these systematic differences and
uncertainties may require further advances in theoretical research,
e.g. in terms of a more detailed and improved understanding of the
late stages of stellar evolution, stellar atmospheric and pulsation
physics, horizontal-branch morphologies, and mass-loss processes,
among others, as a function of stellar mass.

From an observational perspective, the future looks bright across the
entire observable wavelength range. Although much current focus is on
designing ever larger telescopes, the astronomical community must
carefully consider whether the field is best served by having access
to the next-generation of these extremely large telescopes at
optical/near-IR wavelengths and the {\sl SKA} in the radio domain or
if significant progress can still be made with dedicated 2--4 m-class
optical telescopes and upgraded current-generation radio
interferometers. Clearly, although they will have small fields of
view, larger optical and near-IR telescopes will have larger
light-collecting areas and we will, thus, be able to apply current
techniques to objects at greater distances: think of e.g.
eclipsing-binary analysis potentially at Virgo cluster distances,
monitoring Cepheid variables spanning a reasonable period distribution
in Coma cluster galaxies and RR Lyrae variables in both spirals and
ellipticals in the Virgo cluster, thus providing an independent
calibration of SN Ia distances and finally linking the different
stellar-population tracers.

On the other hand, one only has to consider the tremendous success of
surveys with small telescopes, such as the {\sc ogle} and the Sloan
Digitial Sky Survey ({\sc sdss}), to realize that smaller, dedicated
telescopes still have an important role to play in the overall context
of astrophysical distance measurement. After all, in many cases
currently unresolved questions benefit from being allocated
significant amounts of observing time rather than access to the deep
Universe. In this context, the European Southern Observatory's {\sl
  VISTA} telescope (Emerson et al. 2004) will likely play an important
role in e.g. achieving firmer zero points for period--luminosity
relations at near-IR wavelengths by surveying the Magellanic Clouds as
well as the Galactic Centre region and the inner disk through the {\sl
  VISTA} near-IR $YJK_\mathrm{s}$ survey of the Magellanic System
(VMC; Cioni et al. 2008, 2011) and the {\sl VISTA} Variables in the
V\'{\i}a L\'actea (VVV; Minniti et al. 2010) public surveys,
respectively.

Looking beyond the immediate future, many new ground-based
observatories, including the {\sl Large Synoptic Survey Telescope
  (LSST)} and {\sl Pan-STARRS} (Panoramic Survey Telescope \& Rapid
Response System), and space-based missions are currently in the
design, construction or early operations phases, at wavelengths across
the electromagnetic spectrum, from the very-high-frequency X-rays
(e.g. in the context of improving Sunyaev--Zel'dovich-effect
measurements) to low-frequency radio waves. In addition, let me
highlight one of the key forthcoming space-based missions of relevance
to the field of astrophysical distance measurement. The Milky Way's
structure will be characterized to unprecedented levels of accuracy
within a few years of the launch of {\sl Gaia}.

Somewhat further afield, the {\sl JWST} will give us an
unpre\-cedentedly high-resolution mid-IR view of the Universe,
promising e.g. significant reduction of the uncertainties in mid-IR
Cepheid period--luminosity relations (e.g. Madore et al.  2009a,b) and
red-giant-branch-bump validation as a distance indicator (e.g. Valenti
et al. 2004), among others. Observations at IR wavelengths hold
significant promise in relation to improved or alternative methods of
distance determination.

Remarkable and significant progress as regards the accuracy and
robustness of cosmic distances at any scale has been made in the past
few decades. The launch of the {\sl HST} in the early 1990s proved a
pivotal event in reducing the uncertainties in the Hubble constant,
predominantly through carefully calibrated Cepheid-based extragalactic
distances. {\sl WMAP} has allowed determination of the prevailing
cosmological parameters as well as the Hubble constant at high
redshift to unprecedented accuracy and precision, provided that the
cosmological-model-dependent assumptions at the basis of these results
retain their validity as ever more precise and larger-scale
measurements are becoming available. Lower rungs of the distance
ladder have also seen (at least partial) convergence of their absolute
levels through cross calibration with independent methods of distance
determination. Nevertheless, establishing a fully robust distance
ladder -- or, as proposed at the meeting, rather a {\it network of
  distance tracers} -- remains a lofty goal and may, in fact, be but
an unreachable dream, given the significant uncertainties affecting
many of the contributing methods, even the most robust techniques
(cf. the Pleiades controversy).

In an attempt at summarising this vast field, Figure
\ref{distanceladder.fig}(\footnote{available from {\tt
    http://kiaa.pku.edu.cn/{\textasciitilde}grijs/distanceladder.pdf}})
visualises the applicability, distance range, mutual dependences and
robustness of many of the most common methods of distance
determination.

\end{document}